\documentclass[aps,pra,twocolumn,groupedaddress,showpacs,superscriptaddress,amssymb,amsmath]{revtex4-1}
\usepackage{soul}
\usepackage{amsmath}
\usepackage{latexsym}
\usepackage{amssymb}
\usepackage{graphics,epstopdf}
\usepackage{hyperref}
\usepackage{xcolor}
\hypersetup{
    colorlinks = true,
    linkcolor =blue,
	citecolor=blue, 
	urlcolor=blue 
}

\newcommand{\ed}{\end{document}}
\newcommand{\beq}{\begin{equation}}
\newcommand{\eeq}{\end{equation}}

\usepackage{mathtools}
\usepackage{comment}

\begin{document}
\title{Generalised Hydrodynamics description of the  Page curve-like dynamics of a freely  expanding  fermionic gas }
\author{Madhumita Saha}
\email{madhumita.saha@icts.res.in }
\affiliation{International Centre for Theoretical Sciences, Tata Institute of Fundamental Research,
Bangalore 560089, India}
\author{Manas Kulkarni}
\email{manas.kulkarni@icts.res.in }
\affiliation{International Centre for Theoretical Sciences, Tata Institute of Fundamental Research,
Bangalore 560089, India}
\author{Abhishek Dhar}
\email{abhishek.dhar@icts.res.in }
\affiliation{International Centre for Theoretical Sciences, Tata Institute of Fundamental Research,
Bangalore 560089, India}

\begin{abstract}

We consider an analytically tractable model  that exhibits the main features of the Page curve characterizing the evolution of entanglement entropy during evaporation of a black hole.
 Our model is a  gas of non-interacting fermions on a lattice  that is released from a box into the vacuum. More precisely, our Hamiltonian is a tight-binding model with a defect at the junction between the filled box and the vacuum. 
  In addition to the entanglement entropy we consider several other observables, such as the spatial density profile and  current, and show that the semiclassical  approach of generalized hydrodynamics provides a remarkably  accurate description of the quantum dynamics including that of the entanglement entropy at all times. Our hydrodynamic results agree closely with those obtained via 
exact microscopic numerics. 
 We find that the  growth of entanglement is linear and universal, i.e, independent of the details of the defect. The decay  shows  $1/t$ scaling for conformal defect while for non-conformal defects, it is slower. Our study shows the power of  the semiclassical approach and could be relevant for discussions on the resolution of  the black hole information paradox. 
\end{abstract}
\date{\today}
\maketitle

{\it Introduction---} Entanglement is a quantity of immense interest~\cite{eisert2010colloquium,nishioka2018entanglement,review_entanglement_entropy,
entanglement_review_calabrese,entanglement_growth,Page_dynamical} spanning  and connecting different  areas such as quantum information, quantum many body  physics, black hole physics to name a few.  
The entanglement entropy characterizes quantum correlations between two parts of a given system
in a pure state and indicates how far it is from a product form. 
An interesting question is that of the time evolution of the entanglement entropy, in a many body system, between a subsystem and its complement. Concrete results were obtained by Calabrese and Cardy for one dimensional noninteracting systems using path integral methods and also explicit calculations on lattice models~\cite{calabrese2005evolution,calabrese2007entanglement, entanglement_review_calabrese}. A number of subsequent papers have studied this problem for both interacting~\cite{alba2021generalized,alba2017entanglement,ingo_peschel_interacting, entanglement_diffusive} and non-interacting~\cite{bertini2018entanglement,logt_analytical_calebrese,entanglementxy_calebrese,logt_hydrodynamics,ingo_peschel1, Full_counting_entanglement,Full_counting_entanglement2,entanglement_hydrodynamics,entanglement_defect_ingo_peschel,defect_calebrese_hydrodynamics} one-dimensional integrable models and in general one finds  that the entanglement entropy initially increases linearly with time (in some cases logarithmically) and eventually saturates to a value that is consistent with the volume law~\cite{ho2017entanglement,kim2013ballistic}.

There is also  a lot of interest in the setup where there is a possible decay in the entanglement entropy at late times. This is often described by the Page curve that is usually discussed in the context of the evaporation  of a black hole. Page~\cite{page1993information,page2013time} considered the entanglement between the black hole and the radiation starting from the unentangled initial state of just the black hole.   As the black hole radiates,  the effective Hilbert space dimension of the radiation increases and,  assuming maximal entanglement, there will be  a corresponding increase in the entanglement entropy. However, this increase has to stop at some time when the black hole and radiation have the same Hilbert space dimensions. Beyond this time (referred to as the Page time), the entropy has to decrease. 
 This non-monotonic behavior  was not captured in Hawking’s semiclassical calculation~\cite{hawking1975particle} leading to the famous information paradox~\cite{almheiri2021entropy,raju2022lessons,mathur2009information}, see also ~\cite{maes2015no} for a critique.
In fact, such non-monotonic behavior is also typically  not  observed in non-equilibrium studies of  quantum many-body systems after a quench (discussed above~\cite{calabrese2005evolution,calabrese2007entanglement}), where typically the two partitions are of comparable size~\cite{ho2017entanglement,kim2013ballistic}, and the entropy increases monotonically and saturates. 

In this work, we show that by considering a partition consisting of a finite system and an infinite environment, one observes all the essential features of the Page entanglement curve, in particular the initial linear growth and the eventual decay. In two  recent papers, this Page curve time dynamics behaviour has been observed  for open fermionic \cite{kehrein2023page} and bosonic \cite{glatthard2024page} systems. Note that this is distinct from works reproducing the page curve in various free theories where entanglement is studied as a function of subsystem fraction for fixed Gaussian states \cite{yu2023free,bianchi2022volume,bianchi2021page,bhattacharjee2021eigenstate}. 

Our microscopic model is the same as studied in Ref.~\onlinecite{kehrein2023page}, namely an expanding free fermionic gas. 
 Our main finding is that the Page curve behavior of entanglement entropy for this system is very accurately described from the equations of semi-classical generalized hydrodynamics  which can be solved exactly. Our exact solution of the hydrodynamic equations  reveals interesting features of the Page curve decay and also allows us to obtain analytic forms for various physical observables such as density profiles and particle current. We verify our hydrodynamics results from exact numerical computations. 
 {Interestingly, we find that even the average particle current has  hints of the full Page curve} while the particle number fluctuation displays the full form of the Page curve. 
 The entropy obtained from hydrodynamics, is usually referred to as the Yang-Yang entropy and is a ``coarse grained entropy" that counts the number of microstates corresponding to a given phase-space density~\cite{alba2021generalized} (see also ~\cite{pandey2023}). Remarkably, we find that the Yang-Yang entropy evaluated for the system agrees with the microscopic entanglement entropy at all times, while the same computed for its complement (the reservoir) 
 differs from the entanglement entropy after around the Page time and keeps increasing.

\begin{figure}
\includegraphics[width=0.95\columnwidth]{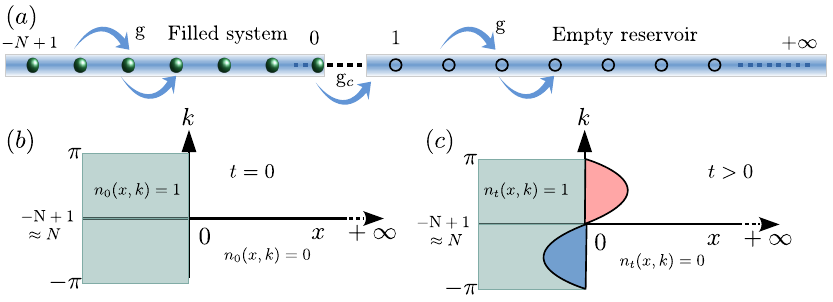} 
\caption{(a) A schematic showing a finite filled fermionic system of length $N$  coupled with an infinite empty fermionic reservoir. The hopping strength for both the finite system and the reservoir is $g$, except at the location of the defect (junction of system-reservoir) where the hopping strength is $g_c$. In (b) is shown the initial phase space density $n_0(x,k)$  while (c) shows the density  $n_t(x,k)$  at a later time $t>0$. Here $x,k$ denote the position and momentum of the particles.} 
\label{schematic} 
\end{figure}

{\it Model---} We consider the set up as shown in Fig.~\eqref{schematic}.
The Hamiltonian for the full non-interacting set up can be written as,
\begin{align}
\label{system}
\hat{H}=\sum_{i,j=-N+1}^{\infty} h_{ij} \hat{c}^{\dagger}_i \hat{c}_{j}.
\end{align}
Here, $\hat{c_i}(\hat{c_i}^{\dagger})$ is the fermionic annihilation (creation) operator at site $i$ of the 1D chain. $h$ is the single particle Hamiltonian with $h_{i,j}=-g (\delta_{i,j+1}+\delta_{i+1,j})\, \forall i,j \neq 1, 0$. Here, $g$ is the nearest neighbor hopping strength except at the coupling between the finite system ($-N+1 \leq i\leq 0$) and the semi-infinite reservoir ($1 \leq i < \infty$). For conformal defect, the coupling matrix elements are given as,
\begin{align}
 h_{0,1}=h_{1,0}=-g_c,~~h_{0,0}=-h_{1,1}=\sqrt{g^2-g_c^2}.
\end{align} 
For other non-conformal defects such as hopping defect and density impurity, the coupling matrix elements of the Hamiltonian are given  as,
\begin{align}
h_{0,1}&=h_{1,0}=-g_c~~~~~\mathrm{for~hopping~defect}, \\ \nonumber
h_{0,0}&=h_{1,1}=g_c~~~~~~~\mathrm{for~density~defect}.
\end{align}
In the limit $N \rightarrow \infty$, the problem reduces to  scattering of plane waves with wave-vector $k$ across the defect with transmission probability $T_k$ and reflection probability $R_k=1-T_k$ given by~\cite{transport_defect}:
\begin{align}
\label{Rk}
R_k&=1-\lambda^2~\quad\rm{for~ conformal~defect}, \\ \nonumber
R_k&=\frac{(\lambda^2-1)^2}{\lambda^4+1-2 \lambda^2 \cos[2k]}~\quad\mathrm{for~ hopping~defect}, \\ \nonumber
R_k&=\frac{\lambda^2(\lambda -2 \cos[k])^2}{2+2 \lambda^2+ \lambda^4-4 \lambda^3 \cos[k] + 2(\lambda^2-1)\cos[2k]} \\ \nonumber
&~~~~\quad\rm{for~density~defect}.
\end{align} 
Here, $\lambda=g_c/g$. For conformal defect, $R_k$ and $T_k$'s are $k$ independent and for non-conformal defect $R_k$ and $T_k$'s are $k$ dependent. Throghout the manuscript, we consider $g=1/2$.

{\it Exact calculations for microscopic entanglement entropy---}
To do the exact numerical calculations, we have to deal with finite-dimensional matrices. Thus, we  write the finite dimensional form of the  Hamiltonian in Eq.~\eqref{system}: 
\begin{align}
\label{system1}
\hat{H}=\sum_{i,j=-N+1}^{N_b} h_{ij} \hat{c}^{\dagger}_i \hat{c}_{j} \equiv D^{\dagger}h D,
\end{align}
where $N_b$ is the length of the reservoir and we take $N_b>>N$. Here $D=\{\hat{c}_i\}$ is a column vector containing all the annihilation operators. Similarly, $D^{\dagger}$ is the row vector containing all the creation operators. The correlation matrix $C(t)$ for any time $t$ can be written as,
\begin{align}
\label{correlation}
C(t)=\langle [D^{\dagger}]^T D^T\rangle.
\end{align}
Here, $T$ symbol is used for the transpose of the matrix. $\langle \hat{c}_i^{\dagger} \hat{c}_j\rangle$ are the matrix elements of correlation matrix $C(t)$. Using the Heisenberg equation of motion, $C(t)$ can be written as,
\begin{align}
\label{solution1}
C(t)=e^{i h t}C(0)e^{-i h t}.
\end{align}
Here, $C(0)$ is the initial correlation matrix. For this work, we  consider the initial condition as filled finite system and empty large (infinite) system i.e. $C(0)_{i,j}=\delta_{i,j} ~\forall -N+1 \leq i,j \leq 0$ and otherwise $0$. Thus the knowledge of the single particle spectrum of $h$ enables us to compute exactly the correlation matrix $C$.
Once we know $C(t)$ from Eq.~\eqref{solution1}, we can evaluate all the physical observables like average current, density, entanglement entropy etc. The average current $ I$ flowing from the  system to the reservoir  can be calculated using the elements of $C(t)$ as, 
\begin{align}
\label{current}
 I = 2 g_c \mathrm{Im}[\langle \hat{c}^{\dagger}_0 \hat{c}_1 \rangle]. 
\end{align}
The average density $\rho(i)$ at any site $i$ and any time $t$ is the diagonal elements of $C(t)$ i.e. 
\begin{align}
\label{density}
\rho(i)=\langle \hat{c}^{\dagger}_i \hat{c}_i \rangle~\forall -N+1\leq i \leq N_b.
\end{align}
The entanglement entropy between the  system and reservoir can be simply obtained~\cite{peschel2009reduced} in terms of the eigenvalues of the correlation matrix, $C_s(t)$, which has the same elements as $C_{ij}$, but restricted to the range $-N+1\leq i,j  \leq 0$. 
 In terms of the eigenvalues, $m_l$, $l=1,2,\ldots, N$ of  $C_s(t)$, the von Neumann entropy $S$ is given by 
\begin{align}
\label{entropy1}
S=-\sum_{l=1}^{N} \big[(1-m_l)\log[1-m_l]+ m_l \log m_l \big].
\end{align}

{\it Particle number fluctuations and entanglement entropy--- }
It has been shown that the entanglement entropy $S$ is directly related to charge statistics in a quantum point contact set up 
as ~\cite{levitov1,Full_counting_entanglement,Full_counting_entanglement2},

\begin{align}
S=\sum_{m>0} \frac{\alpha_m}{m!} \kappa_m,\quad \alpha_m =\begin{cases} (2\pi)^m |B_m| \quad \text{$m$ even} \\ 0 \quad \text{$m$ odd} \end{cases}
\end{align}
where $\kappa_m$ are the cumulants corresponding to particle number fluctuations in the system and  $B_m$ are Bernoulli numbers.

Thus, the major contribution to the entropy comes from the 
second cumulant $\kappa_2$ which, in terms of eigenvalues of $C_s(t)$, $\kappa_2$ can be written as ~\cite{Full_counting3}, 
\begin{equation}
\label{eq:kappa}
\kappa_2=\sum_{\ell=1}^{N} m_{\ell}(1-m_{\ell}).
\end{equation}
\begin{figure*}
\includegraphics[width=16cm]{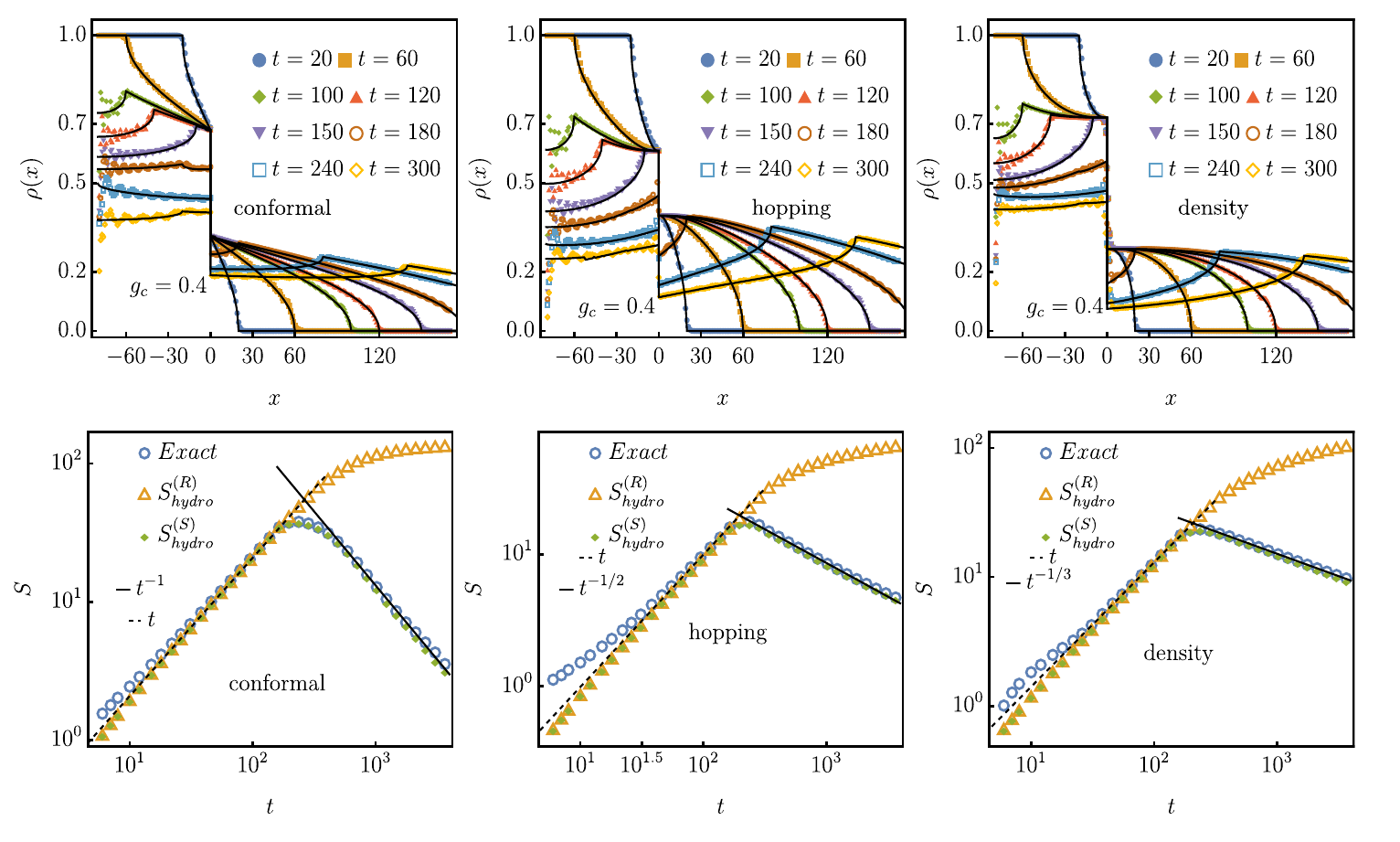} 
\caption{(Top panel) Plot of the density profiles at different times for the three different types of defect. 
We show  results both from exact   numerics (colored lines)  and from the hydrodynamic approach (black lines). We  see that the hydrodynamic approach agrees well with the exact numerics for all time $t$.  It fails to capture some oscillations seen in the  exact numerics.
(Bottom panel) Entropy as a function of time. We also see that the hydrodynamic system entropy $S_{hydro}^{(S)}$ agrees with the microscopic entanglement entropy $S$ for almost all time $t$. On the other hand, the hydrodynamic reservoir entropy $S_{hydro}^{(R)}$ starts differing beyond the Page time $t_P \sim 2N$. It deviates from the Page curve and appears to keep growing.  The entanglement entropy increases  linearly with time  till up to around  $t_P$, and then has a power-law decay $1/t^a$, with $a=1,1/2,1/3$ for the conformal, hopping and density defects respectively. For the case of hopping defect, we eventually expect $a=1/3$ (see Appendix \ref{app:above_pt}) similar to to density defect in the very long time limit. For all the plots, the system size was $N=80$, reservoir size $N_B=4096$, and $g_c=0.4$.} 
\label{fig:main} 
\end{figure*}
Next, we will see, how using the generalized hydrodynamic description, we can calculate the quantities like average current, density and hydrodynamic entanglement entropy for this set up. Note that, the particle number fluctuation ($\kappa_2$) has been computed~\cite{Full_counting_entanglement} from hydrodynamics for the case of infinite $N$ but we are not aware of results for finite $N$. 

{\it Hydrodynamic description---}  The evolution of integrable systems observed on large time and length scales is described by generalized hydrodynamics~\cite{castro2016emergent,bertini2016transport,doyon2020lecture,bertini2018entanglement,essler2022short}  which views the system as a gas of quasiparticles which carry fixed momentum labels $k$ and have a  phase space density $n_t(x,k)$, with $x \equiv i$.  For non-interacting systems such as the system of free fermions considered here, the quasiparticle velocities are given by $v_k=\sin[k]$, and the evolution of $n_t(x,k)$ is given by the Euler equation,
\begin{align}
\label{Euler_equation}
\big(\partial_t + \sin[k] \partial_x\big) n_t(x,k)=0.
\end{align}
This equation has to be solved with appropriate boundary conditions, which for our set-up are: (a) left-moving quasiparticles are reflected at the boundary $x=-N$ (as $-N+1\approx N$); (b) at the defect, right-moving  quasiparticles inside the system are reflected  (transmitted) with probability $R_k$ ($T_k)$, given by Eqs.~\eqref{Rk}.
In terms of the quasiparticle distribution, physical observables such as the density profile  and the hydrodynamic entropy density are given by 
\begin{align}
\label{density_hydro}
\rho(x)&=\int_{-\pi}^\pi \frac{dk}{2\pi} n_t(x,k), \\
s_{hydro}(x)&=- \int_{-\pi}^\pi \frac{dk}{2\pi} ~\Big[n_t(x,k) \log(n_t(x,k)  \nonumber\\
   &+\big(1-n_t(x,k)\big) \log \big(1-n_t(x,k)\big) \Big]. \label{entropydens-hydro}
\end{align}
This is the thermodynamic entropy density and also 
referred to as the Yang-Yang entropy ~\cite{yang-yang-entropy,yang-yang-entropy2}. The system and reservoir entropies are then  given by,
\begin{align}
S_{\rm hydro}^{(S)}&= \int_{-N}^0 \, dx \, s_{\rm hydro}(x) \label{entropyS-hydro}, \\ 
S_{\rm hydro}^{(R)}&= \int_{0}^\infty dx \, s_{\rm hydro}(x) \label{entropyR-hydro}.
\end{align}
Unlike for the case of  domain wall initial conditions studied earlier in the literature~\cite{defect_calebrese_hydrodynamics}, our set up lacks particle-hole symmetry which means that $S_{\rm hydro}^{(S)}$ and $S_{\rm hydro}^{(R)}$ are not equal at all times. As we will see, it is $S_{\rm hydro}^{(S)}$ that in fact gives precisely the entanglement entropy. Other observables, such as the total particle number in the system  is simply given by $\mathcal{N}=\int_{-N}^0 dx \rho(x)$,  while the current into the reservoir is $I=-d \mathcal{N}/dt$. 

% \mk{is not necessary that both ways of calculating entanglement entropy will give identical results. In other words, there is no particle-hole symmetry in our setup.} 

We now proceed to obtain a solution of Eq.~\eqref{Euler_equation} with the boundary conditions at the bounding wall at $x=-N$ and at the defect at $x=0$. 
We first note that the solution on the infinite line is given by 
\begin{equation}
n_t(x,k)=n_0(x-t \sin[k],k).
\end{equation}
Here the initial phase space density, corresponding to  the fully filled lattice is given by
\begin{align}
\label{initial1a}
n_0(x,k)=\theta(-x)-\theta(-x-N)\, \quad \text{for}\,\, -\pi\leq k \leq \pi.
\end{align}

In the presence of the defect at $x=0$, the phase space density for $x>0$ is the sum of contributions where the final momentum of the quasi-particles is $k>0$. Such contributions can occur in two ways  (i) an initially right moving quasi-particle with momentum $k>0$ gets transmitted from $x<0$ to $x>0$ directly or after having multiple reflections due to the defect and boundary of the finite system, (ii) an initially left moving quasi-particle with momentum $k<0$ gets reflected once or multiple times from the boundary and the defect and then reaches $x>0$ after one transmission. Considering both the contribution, we can write $n_t(x>0,k)=n_t(x>0,k>0)$ as,
\begin{align}
\label{phase_spacexg0}
n_t(x\!>0, k>0)= \sum_{s=0}^{\infty} T_k R_k^s \Big(n_0(x+2 s N-t \sin[k]) \nonumber \\+ 
n_0(-x-2 s N -2N+t \sin[k]) \Big).
\end{align}
Here $s$ is the number of reflections. 
% Applying the initial condition given in Eq.~\eqref{initial1a} in Eq.~\eqref{phase_spacexg0} and after some simplifications, $n_t(x>0,k)$ can be written as,
% \begin{align} \label{final_phase_space_xg0} n_t(x>0,k)=\sum_s T_k R_k^s \Big(\theta(-x-2 s N+t \sin[k],k)\\ \nonumber -\theta(-x-2 s N-2N+t \sin[k],k) \Big).
% \end{align}
Similarly for $x<0$, we can calculate phase space density where both final momentum $k<0$ and $k>0$ will contribute to the phase space density. Thus, four possible contributions for $x<0$ regime are (i) initially quasi particles with $+k$ momentum will be at at $x<0$ with final $+k$ momentum (ii) initially quasi particles with $+k$ momentum will be at at $x<0$ with final $-k$ momentum after one or multiple reflections (iii) initially quasi particles with $-k$ momentum will be at at $x<0$ with final $-k$ momentum (iv) initially quasi particles with $-k$ momentum will be at at $x<0$ with final $+k$ momentum after one or multiple reflections. Adding all these contributions we can write down the phase space density $n_t(x<0,k)$ as 
\begin{equation}
n_t(x<0,k)=n_t(x<0,k>0)+n_t(x<0,k<0)\, ,
\end{equation}
with 
\begin{align}
\label{phase-spacexl0}
n_t(x<0,k>0)&= \sum_{s=0}^{\infty}  R_k^s \Big(n_0(x+2 s N-t \sin[k])  \nonumber\\
&+ n_0(-x-2 s N -2N+t \sin[k])\Big), \\ \nonumber
n_t(x<0,k<0)&= \sum_{s=0}^{\infty} R_k^{s+1} n_0(-x+2s N+t \sin[k]) \\ 
&+ R_k^s n_0(x-2s N -t \sin[k]). \nonumber
\end{align} 
Equations \eqref{phase_spacexg0}, \eqref{phase-spacexl0} along with Eq.~\eqref{initial1a} provide a complete explicit solution for the phase space density at all times. Various asymptotic results, namely at short  and late times, can be obtained in more explicit forms and are given in the Appendix \ref{appendixa}. One main observation is that time-dependence undergoes a drastic change  at the ``Page" time $t_P = 2N/v_F$, where for our case, the Fermi velocity  $v_F=1$. 
Here we summarize some of our results on the form of the entropy and current at times $t\lesssim t_P$ and $t >> t_P$: 
\begin{align}
\label{hydro-asymp}
 S &\sim \begin{cases}\alpha \,t \quad~~~~{\rm for}~~~  t \lesssim t_P 
 , \\
1/t^a~~~~{\rm for}~~~ t >> t_P \end{cases}\\
I & \sim \begin{cases}\beta  ~~~~~~~~{\rm for}~~~ t \lesssim t_P, \\  1/t^{a+1} ~{\rm for}~~~ t >> t_P \end{cases}\\ \nonumber
\end{align}
Here the value of $\alpha,\beta$ depends on the type of defect (see Appendix \ref{app:before_pt}) and the exponent $a$ takes the values $1,1/3,1/3$ for conformal, hopping and density defect respectively. Thus we observe the proportionality $|dS/dt| \propto I$ at both early and late times.

Next, we present results for current, density and entanglement entropy from exact microscopic calculations and compare them with the  hydrodynamic description.\\

{\it Comparison between exact numerics and hydrodynamics---} For the exact numerics, the average current $I$, density $\rho(i)$ or $\rho(x)$, number fluctuations $\kappa_2$ and the entanglement entropy $S$ can be calculated using Eqs.~[\eqref{current},\eqref{density},\eqref{eq:kappa},\eqref{entropy1}], while from  hydrodynamics, we can calculate these quantities using Eqs.~[\eqref{density_hydro},\eqref{entropydens-hydro},\eqref{entropyS-hydro},\eqref{phase_spacexg0},\eqref{phase-spacexl0}]. In Fig.~\eqref{fig:main}, the top panel shows  the  density profiles at different times for all three types of defects.  We see excellent agreement between exact numerics and hydrodynamics  at all times, though we note that hydrodynamics fails to capture the oscillations seen in the density profile from exact numerics. The profiles at early times $t < N$ are the same as obtained in earlier studies for infinite line domain wall evolution~\cite{transport_defect,defect_calebrese_hydrodynamics}. After $t> 2N$, which is the time for the fastest particle with speed $v_F=1$ to make a round trip across the system, the effect of the finite system size shows up in the entire density profile. 
 In the lower panel in Fig.~\eqref{fig:main}, we have plotted the microscopic entanglement entropy from exact numerics  and the coarse-grained hydrodynamic  entropies computed for system, i.e $S^{(S)}_{\rm hydro}$,  and for the reservoir, i.e $S^{(R)}_{\rm hydro}$.  We see that  at times $t \lesssim  2N$  all three entropies agree and we see a linear growth with time. Beyond this time, the entanglement entropy shows a decay, as expected from the Page curve --- we thus refer to $t_P=2N$ as the Page time. Remarkably, beyond the  Page time, the    entanglement entropy is perfectly reproduced  by the system hydrodynamic entropy $S^{(S)}_{\rm hydro}$, both of which show a decay. On the other hand the reservoir hydrodynamic entropy, $S^{(R)}_{\rm hydro}$, shows a monotonic growth beyond the Page time.   The long time decay  of $S$ for all three different defects has the form $1/t^{a}$, expected from hydrodynamics, see Eq.~\eqref{hydro-asymp}. Finite size effects can also be understood completely from the hydrodynamic theory and we show some numerical verifications in Appendix~\ref{app:finite}. In particular we find the scaling form $S=N F(t/N)$ where $F$ is a scaling function.
 \begin{figure}
\includegraphics[width=\columnwidth]{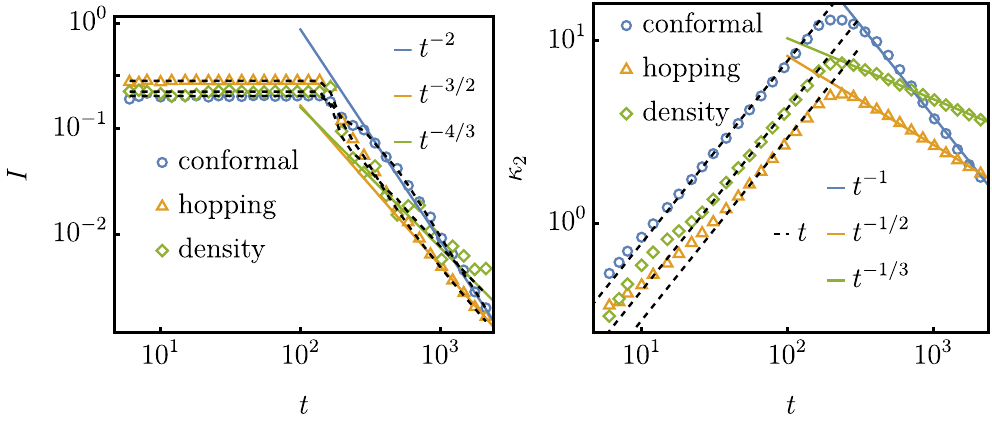} 
\caption{ (a) Plot of the time dependence of current, $I$ for the three different defects. The current is flat till the Page time $t\sim 2N$ and then shows a power law decay regime. 
We again see very good agreement with hydrodynamics (black dashed lines). For the hopping defect, we expect to get $t^{-4/3}$ in the very long time limit (see Appendix \ref{app:above_pt}).  (b) particle number fluctuations, $\kappa_2$ as a function of time. We  see that $\kappa_2$ shows the Page-curve-like dynamics with initially linear growth. The decay of $\kappa_2$ with time $t$ also captures  behavior similar to that of entanglement entropy.}
\label{fig:number} 
\end{figure}

In Fig.~\eqref{fig:number} we show plots of the dynamics of 
 average current, $I$, and the number fluctuations $\kappa_2$, again for the three different defect types. Up to $t^* \sim t_P$, the current is almost constant for all the three defects and the value of the constant current is governed by the transmission probabilities. At long times we again see the  decay $t^{-(a+1)}$ expected from hydrodynamics (Eq.~\eqref{hydro-asymp}). Overall, the current behavior is consistent with the form $|dS/dt \propto I$ at all times. 
The number fluctuations $\kappa_2$ closely follows the form of $S$ and  shows the Page curve like form with identical short and long time scalings. It is worth noting that $\kappa_2$ is a more experimentally accessible quantity than entanglement entropy as mentioned in ~\cite{levitov1} and thus finding the identical Page curve like dynamics for $\kappa_2$ is interesting and of relevance.

{\it Summary and outlook ---} 
To summarize, we considered a freely expanding fermionic  gas for which the detailed dynamical structure of the Page curve for entanglement was elucidiated through exact numerics and analytic calculations based on semiclassical generalized hydrodynamics. 
 We investigated spatial density profiles, current and entanglement entropy and show that  generalized hydrodynamics (semiclassical) provides a remarkably accurate description of the microscopic quantum dynamics. In this setup, a defect at the system-reservoir interface is crucial for generating entanglement and we studied different defect types. We showed that the  early time growth of entanglement is linear, independent of the defect type, while the long time decay depends on the nature of the defect.
Interestingly, we observe that  the system hydrodynamic entropy captures the full microscopic Page curve while the reservoir hydrodynamic entropy  agrees with the Page curve only before the Page time. Beyond this time it appears to grow monotonically.   
 Our approach could be of relevance for further studies on Hawking's semiclassical calculation and the black hole information paradox.

 {%Deeper physical reasons on why the hydrodynamic entropy calculated from system agrees remarkably with the microscopic von Neumann entropy is a fascinating challenging question. 
 It will be interesting to study dynamics with other initial conditions such as partial filling fraction, finite temperatures, and to explore the applicability of generalized hydrodynamics (or lack thereof) in such cases. Even in the absence of defects, entanglement entropy for the similar set up shows a logarithmic increase with time and then a decay for a finite system. The logarithmic increase cannot be captured by generalized hydrodynamics and requires one to  consider  quantum fluctuations~\cite{logt_hydrodynamics}. Understanding the full growth and subsequent decay of entanglement entropy for the defect-free case using quantum fluctuating hydrodynamics will be interesting. 
 Exploring Page curve-like dynamics for interacting cases, possibly toy models of black holes, and  their hydrodynamic description are other fascinating questions.}\\

{\it Acknowledgements ---}
AD thanks Raghu Mahajan for very useful discussions. MS would like to thank Archak Purkayastha for useful discussions. 
MK and AD acknowledge support from the Department of Atomic Energy, Government of India, under Project No. RTI4001. MK and AD thank the VAJRA faculty scheme (No. VJR/2019/000079) from the Science and Engineering Research Board (SERB), Department of Science and Technology, Government of India. AD acknowledges the J.C. Bose Fellowship (JCB/2022/000014) of the Science and Engineering Research Board of the Department of Science and Technology, Government of India.

\appendix

\setcounter{figure}{0}
\renewcommand{\thefigure}{A\arabic{figure}}

\section{Analytical long time and short time behavior of current, hydrodynamic entropy for different defects}
\label{appendixa}
Using phase space density $n_t(x,k)$ from hydrodynamic description, it is possible to calculate density $\rho(x)$, average current $I$ and hydrodynamic entanglement entropy $S_{hydro}^{(S)}$ or $S_{hydro}^{(R)}$. The formula for phase space density for $x>0$ is given as, 
\begin{align}
\label{final_phase_spacexg0}
n_t(x>0,k)&=n_t(x>0,k>0); \\ \nonumber
&=\sum_{s=0}^{\infty} T_k R_k^s \big(\theta(-x-2 s N+t \sin[k])\\ \nonumber
&-\theta(-x-2 s N-2N+t \sin[k])\big). 
\end{align}
For $x<0$ we get, 
\begin{align}
\label{app_totaln}
n_t(x<0,k)&=n_t(x<0,k>0)+n_t(x<0,k<0); 
\end{align}
where,
\begin{align}
\label{final_phase_spacexl0}
n_t(x<0,k>0)&=\sum_{s=0}^{\infty}  R_k^s \big(\theta(-x-2 s N+t \sin[k])  \nonumber\\
&- \theta(-x-2 s N-2N+t \sin[k])\big), \nonumber \\
n_t(x<0,k<0)&=\sum_{s=0}^{\infty} R_k^{s+1} \big(\theta(x-2s N-t \sin[k])  \nonumber \\
&-\theta(x-2s N-N-t \sin[k])\big)  \nonumber \\
&+R_k^s \big(\theta(-x+2s N +t \sin[k]) \nonumber \\
&-\theta(-x+2s N-N +t \sin[k])\big).
\end{align}
For conformal defect, where $R_k=1-\lambda^2$ and $T_k=\lambda^2$ are independent of $k$, density $\rho(x)$ can be calculated analytically using Eq.~\eqref{density_hydro} for all time $t$ as,
\begin{align}
\label{conformal_density}
  \rho(x>0)= \frac{\lambda^2}{\pi} \bigg{[}\sum_{s=0}^{\big[\frac{t-x}{2N}\big]}(1-\lambda^2)^s \cos^{-1}\big[\frac{x+2s N}{t}\big]- \\ \nonumber
  \sum_{s=0}^{\big[\frac{t-x}{2N}-1\big]}(1-\lambda^2)^s \cos^{-1}\big[\frac{x+2s N+2N}{t}\big]\bigg{]}, \\ \nonumber
  \rho(x<0)= \frac{1}{\pi}\bigg{[}\sum_{s=0}^{[\frac{t-x}{2N}]}(1-\lambda^2)^s \cos^{-1}\big[\frac{x+2s N}{t}\big] \\ \nonumber
  - \sum_{s=0}^{\big[\frac{t-x}{2N}-1\big]}(1-\lambda^2)^s \cos^{-1}\big[\frac{x+2s N+2N}{t}\big]\bigg{]} \\ \nonumber
  + \frac{1}{\pi}\bigg{[}\sum_{s=0}^{\big[\frac{|t+x|}{2N}\big]}(1-\lambda^2)^{s+1} \cos^{-1}\big[\frac{2s N-x}{t}\big] \\ \nonumber
  -\sum_{s=0}^{\big[\frac{|t+x|}{2N}-1\big]}(1-\lambda^2)^{s+1} \cos^{-1}\big[\frac{2s N+2N-x}{t}\big]\bigg{]}\, , 
\end{align}
where the symbol $[z]$ in the summation index stands for Floor function, i.e., largest integer less than or equal to $z$.
%For all the conformal and non-conformal defects, we can calculate the features of current, density and hydrodynamic entropy before the Page time $t_P$ analytically. We discuss this in the following subsection.
Next we discuss details of density, current  and hydrodynamic entropy for all defects before the Page time $t_P$.

\subsection{Density, current, entropy before the Page time ($t<<t_P$)}
\label{app:before_pt}

In this subsection, we will discuss the case where the time is below the Page time. 

{\it Phase space density:} To see the behavior before the Page time, we can set $s=0$ and 
$N \rightarrow\infty$ in the phase space density given in Eq.~\eqref{final_phase_spacexg0} and Eq.~\eqref{final_phase_spacexl0}. For $x>0$, the phase space density can be written as,
\begin{align}
\label{before_page_xg0}
&n_t(x>0,k)=n_t(x>0,k>0)=T_k \theta(-x+t \sin[k]), 
\end{align}
while for $x<0$, it is
\begin{align}
\label{app_xl0}
n_t(x<0,k)=n_t(x<0,k>0)+n_t(x<0,k<0)
\end{align}
with
\begin{align}
n_t(x<0,k>0) &=  \theta(-x+t \sin[k]), \\ \nonumber
n_t(x<0,k<0) &= R_k (\theta(x-t \sin[k]) + \theta(-x+t \sin[k]).
\end{align}
{\it Density:}
Using Eq.~\eqref{before_page_xg0} and Eq.~\eqref{app_xl0}, we can write the formula for density $\rho(x)$ for both $x>0$ and $x<0$ as,
\begin{align}
\label{rhobeforepage}
\rho(x>0)&=\int_{0}^{\pi}T_k \theta(-x+t \sin[k]) \frac{dk}{2\pi} \\ \nonumber
&= \frac{1}{2\pi}\int_{\sin^{-1}[\frac{x}{t}]}^{\pi-\sin^{-1}[\frac{x}{t}]}T_k  dk, \\ \nonumber
\rho(x<0)&=\int_{0}^{\pi}  \theta(-x+t \sin[k]) \frac{dk}{2\pi} \\ \nonumber
&+ \int_{-\pi}^{0} R_k  (\theta(x-t \sin[k]) \frac{dk}{2\pi} \\ \nonumber
 &+ \int_{-\pi}^{0} \theta(-x+t \sin[k]) \frac{dk}{2\pi} \\ \nonumber
&=\frac{1}{2} + \int_{\sin^{-1}[-\frac{x}{t}]}^{\pi-\sin^{-1}[-\frac{x}{t}]} R_k  \frac{dk}{2 \pi}+ \frac{1}{\pi} \sin^{-1}\Big[-\frac{x}{t}\Big] \\ \nonumber
&=1-\frac{1}{2 \pi}\int_{\sin^{-1}[-\frac{x}{t}]}^{\pi-\sin^{-1}[-\frac{x}{t}]}T_k \,dk.
\end{align}
\noindent
Conformal defect: For conformal defect, $R_k=1-\lambda^2$ and $T_k=\lambda^2$ using Eq.~\eqref{rhobeforepage}, we find 
\begin{align}
\label{rhobeforepage_conformal}
\rho(x>0)&=\frac{\lambda^2}{\pi} \cos^{-1} \Big[\frac{x}{t} \Big], \\ \nonumber
\rho(x<0)&=1-\frac{\lambda^2}{\pi} \cos^{-1}\Big[\frac{|x|}{t}\Big].
\end{align}
%This completely agrees with Eq.~\eqref{conformal_density}.

\noindent
Hopping defect: For hopping defect [Eq.~\ref{Rk}], we get,

\begin{align}
\label{hopping_before_page}
\rho(x>0) &= \frac{1}{\pi} \cos^{-1}\big[\frac{x}{t}\big] 
-\frac{1-\lambda^2}{1+\lambda^2}\bigg[\frac{1}{2} \nonumber \\
&+ \frac{1}{\pi} \tan^{-1}\big[-\frac{(1+\lambda^2)x}{(1-\lambda^2)t \sqrt{1-x^2/t^2}}\big]\bigg], \nonumber \\
\rho(x<0) &= 1-\frac{1}{\pi} \cos^{-1}\big[\frac{|x|}{t}\big] 
+ \frac{1-\lambda^2}{1+\lambda^2}\bigg[\frac{1}{2} \nonumber \\
&+ \frac{1}{\pi} \tan^{-1}\big[-\frac{(1+\lambda^2)|x|}{(1-\lambda^2)t \sqrt{1-|x|^2/t^2}}\big]\bigg].
\end{align}
Note that putting $\lambda=1$, in Eq.~\eqref{hopping_before_page}, gives us the density profile in the defectless case. 

\noindent
Density defect: In the case of density defect, 
% \begin{align}
% R_k=\frac{\lambda^2(\lambda -2 \cos[k])^2}{2+2 \lambda^2+ \lambda^4-4 \lambda^3 \cos[k] + 2(\lambda^2-1)\cos[2k]}. 
% \end{align}
the transmission probability $T_k=1-R_k$ [see Eq.~(\ref{Rk})]  is not symmetric about $\pi/2$. The maximum value of $T_k$ corresponds to the $k=\sin^{-1}[\sqrt{1-\lambda^2/4}]$ which is not $\pi/2$. The expressions for $\rho(x>0)$ and $\rho(x<0)$ are given by,
\begin{widetext}
\begin{align}
\label{density_defect1}
\rho(x>0)&=\frac{1}{2\pi(\lambda^4-3\lambda^2+2)}\bigg[-\big((\lambda^2-2)\pi \big)+ 2(\lambda^2-2) 
\sin^{-1}\Big[\frac{x}{t}\Big]+ \lambda(2+\lambda-\lambda^2) \tan^{-1}\Big[\frac{(\lambda-2)\cot\big[\sin^{-1}[x/t]/2\big]}{\lambda}\Big]+ \\ \nonumber
&\lambda(\lambda^2+\lambda-2) \tan^{-1}\Big[\frac{\lambda\cot[\sin^{-1}[x/t]/2]}{2+\lambda}\Big]+ 
\lambda(\lambda+1)(\lambda-2) \tan^{-1}\Big[\frac{(\lambda-2)\tan[\sin^{-1}[x/t]/2]}{\lambda}\Big]- \\ \nonumber
&\lambda(\lambda+2)(\lambda-1) \tan^{-1}\Big[\frac{\lambda\tan\big[\sin^{-1}[x/t]/2]}{\lambda+2}\Big]\bigg], \\ \nonumber
\rho(x<0)&=1-\frac{1}{2\pi(\lambda^4-3\lambda^2+2)}\bigg[-\big((\lambda^2-2)\pi\big)+ 2(\lambda^2-2) 
\sin^{-1}\Big[\frac{|x|}{t}\Big]+ \\ \nonumber
&\lambda(2+\lambda-\lambda^2) \tan^{-1}\Big[\frac{(\lambda-2)\cot[\sin^{-1}[|x|/t]/2]}{\lambda}\Big]+ 
\lambda(\lambda^2+\lambda-2) \tan^{-1}\Big[\frac{\lambda\cot[\sin^{-1}[|x|/t]/2]}{2+\lambda}\Big]+ \\ \nonumber
&\lambda(\lambda+1)(\lambda-2) \tan^{-1}\Big[\frac{(\lambda-2)\tan[\sin^{-1}[|x|/t]/2]}{\lambda}\Big]- 
\lambda(\lambda+2)(\lambda-1) \tan^{-1}\Big[\frac{\lambda\tan[\sin^{-1}[|x|/t]/2]}{\lambda+2}\Big]\bigg].
\end{align}
\end{widetext}
Again, the special case $\lambda=0$ in Eq.~\eqref{density_defect1}, gives the densities in absence of any defect.

{\it Current:} Using the expression for density in Eq.~\eqref{rhobeforepage} and its continuity equation $\partial_t\rho (x,t)+\partial_x j (x,t)=0$, the current is given by $I = j(0,t)$.
For our set-up, before the Page time, the formula for current can be written as,
\begin{align}
\label{current1}
I=\int_{0}^{\pi} T_k \sin[k] \frac{dk}{2\pi}.
\end{align}
We now give the analytical expressions for various kinds of defect. \\ 
Conformal defect: Using Eq.~\eqref{current1}, we get $I=\lambda^2/\pi$.

Hopping defect: For hopping defect, the analytical form of current is
\begin{align}
I=\frac{1}{\pi}-\frac{(\lambda^2-1)^2 \tanh^{-1}[\frac{2\lambda}{1+\lambda^2}]}{2\pi (\lambda^3+1)}.
\end{align}

Density defect: For density defect, the analytical form of current is,
\begin{align}
I =&\frac{1}{\pi (1-\lambda^2)}+ \frac{\lambda^2 (\lambda-2)^2 \log[(\lambda-2)^2]}{8 \pi (\lambda-1)^2 (\lambda^2-2)} \\ \nonumber
-&\frac{\lambda^2 (\lambda^4-3\lambda^2+4)\log[\lambda]}{2\pi (\lambda^2-1)^2(\lambda^2-1)}+\frac{\lambda^2 (2+\lambda)^2\log[2+\lambda]}{4\pi (\lambda+1)^2(\lambda^2-2)}.
\end{align}
\noindent
\textit{Hydrodynamic entropy:} Before the Page time, calculating hydrodynamic Yang-Yang entropy from both the system's phase space density and reservoir's phase space density give identical result. Thus, here we calculate $S_{hydro}^{(R)}$. Using Eq.~\eqref{rhobeforepage}, we can write the formula for $S_{hydro}^{(R)}$ as,

\begin{align}
\label{entropy_before_page}
S_{hydro}^{(R)}&=-\int_{0}^{t} \int_{0}^{\pi} dx \bigg[ T_k \theta(-x+t \sin[k]) \nonumber \\
&\log\big[T_k \theta(-x+t \sin[k])\big] +(1-T_k \theta(-x+t \sin[k])) \nonumber \\
& \log\big[(1-T_k \theta(-x+t \sin[k]))\big]  \bigg]\frac{dk}{2\pi}.
\end{align}

Eq.~\eqref{entropy_before_page} further simplifies to 
\begin{align}
\label{simp_entropy_before_page}
S_{hydro}^{(R)}=-t \int_{0}^{\pi} \frac{dk}{2\pi} \sin(k) \Big[ T_k  \log(T_k) \\ \nonumber
+(1-T_k) \log(1-T_k )\Big]\, .
\end{align}
Therefore, for any defect the early time entropy is always linear in $t$. The coefficient of the linear term [see Eq.~\eqref{simp_entropy_before_page}] depends on the nature of the defect. In the case of conformal defect, Eq.~\eqref{simp_entropy_before_page} can be calculated exactly and the analytical form is given by,
\begin{align}
S_{hydro}^{(R)}= 
-\frac{(\lambda^2 \log [\lambda^2]+ (1-\lambda^2) \log [1-\lambda^2])} {\pi}~t.
\end{align}
In principle, before the Page time, one can calculate the analytical expressions  $S_{hydro}^{(R)}$ for other defects also. We have presented only the conformal defect here for the sake of brevity. 

\begin{figure}
\includegraphics[width=9cm]{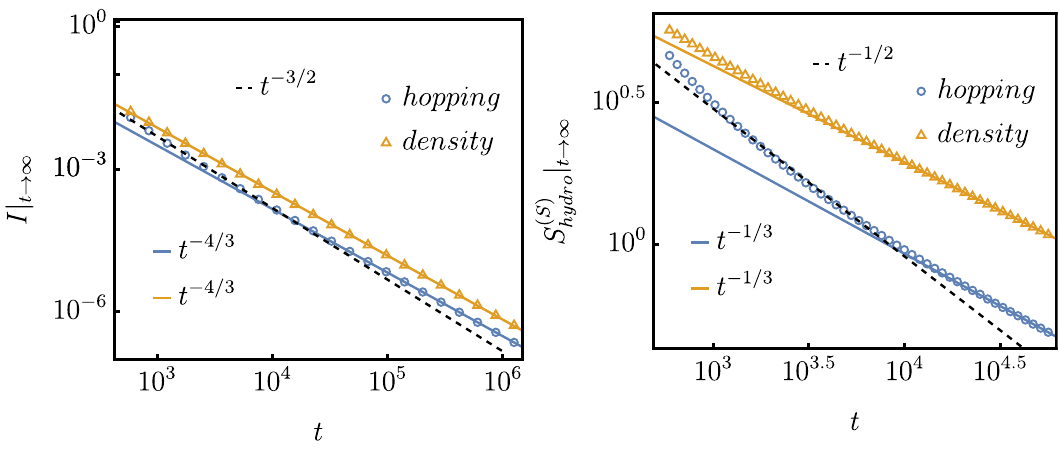} 
\caption{Here, we have calculated the long-time asymptotic  of current $I$ and hydrodynamic entropy $S_{hydro}^{(S)}$ using Poisson summed phase space density for non-conformal defects. For density defect, we can see, current and hydrodynamic entropy show $t^{-4/3}$ and $t^{-1/3}$ scaling after the Page time. However, for hopping defect, there is an intermediate time scale just after the Page time where $I$ and $S_{hydro}^{(S)}$ show $t^{-3/2}$ and $t^{-1/2}$ scaling respectively and then eventually show a crossover to $t^{-4/3}$ and $t^{-1/3}$ scaling .} 
\label{fig:appendix_other_defect} 
\end{figure}

\begin{figure*}
\includegraphics[width=16cm]{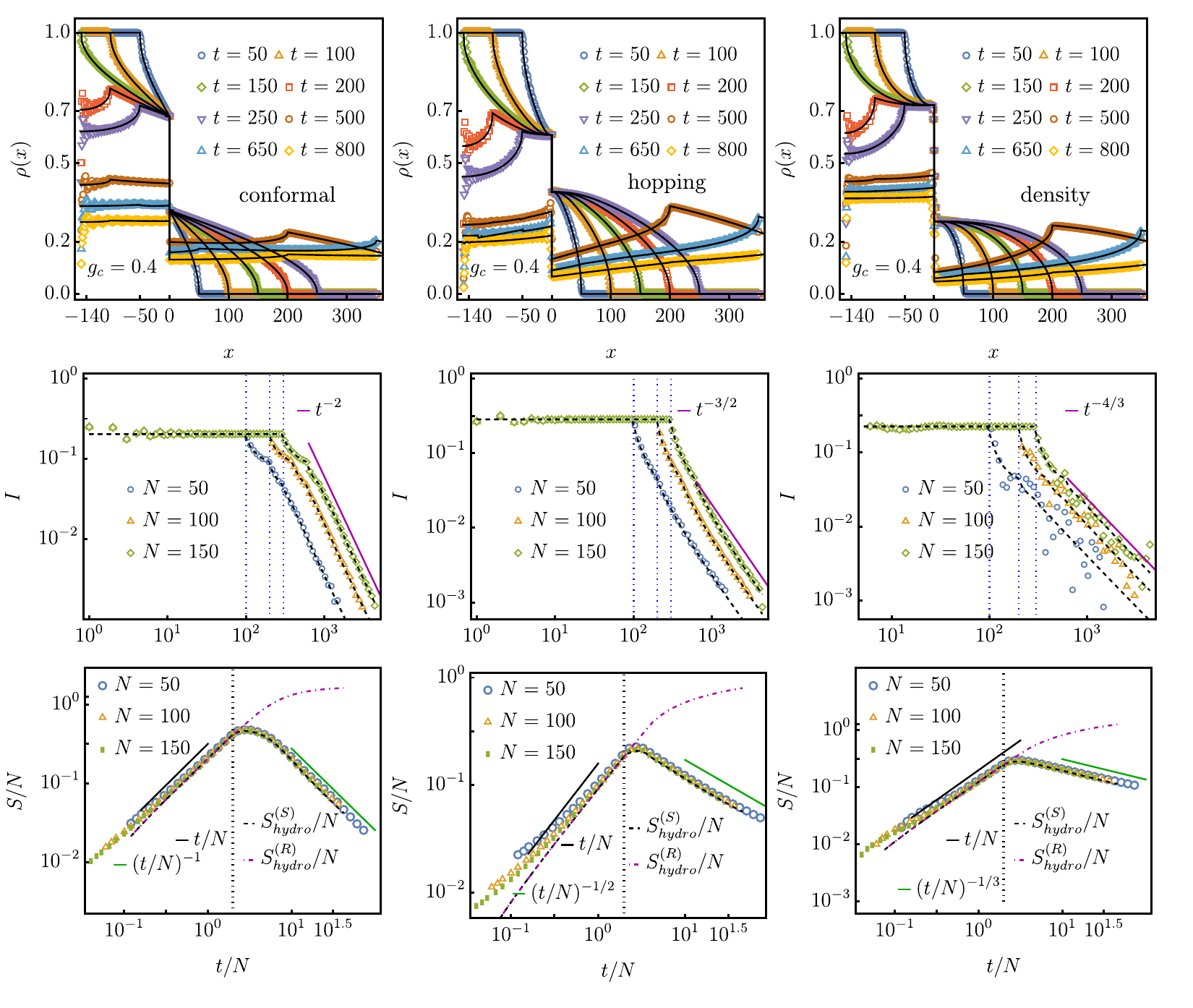} 
\caption{(Finite size effects: Top panel) we have plotted density $\rho(x)$ for different time $t$ for three different defects such as conformal, hopping and density defects for system size $N=150$ from exact numerics (colored lines) and hydrodynamic description (black lines). We can again see that hydrodynamics gives remarkable agreement for density profiles. (Middle panel) We have plotted the current $I$ with time $t$ for three different defects for system size $N=50,100$ and $150$ respectively. The vertical dotted lines are drawn to show the Page time $t_P=2N$. Before the Page time, the current $I$ is constant. After the Page time, for conformal, hopping and density defects, current shows a power-law decay $t^{-2}$, $t^{-3/2}$ and $t^{-4/3}$ respectively. Though from the asymptotic analysis of hydrodynamic phase space density, it is clear that for the hopping defect, current eventually shows a crossover to $t^{-4/3}$ scaling. The black dashed line corresponds to  hydrodynamic results. (Bottom panel) Here we have plotted, entanglement entropy $S/N$ with $t/N$ for three different defects for three different system sizes $N=50,100$ and $150$. The vertical dotted line is drawn near the Page time. We have again shown the linear growth and decay after Page time for all the three defects. For conformal, hopping and density defects, the decay scalings are $t^{-1}$, $t^{-1/2}$ and $t^{-1/3}$. But, eventually for the hopping defect also, the decay scaling is $t^{-1/3}$ which we have seen from asymptotic analysis. $S_{hydro}^{(S)}$ remarkably matches with the microscopic von Neumann entropy for all times, all system size $N$ and all defects. } 
\label{fig:main_appendix} 
\end{figure*}

\subsection{Density, current, entropy after the Page time ($t>>t_P$)}
\label{app:above_pt}

In this subsection, we will discuss the case where the time is larger than the Page time. 

Following the Poisson summation, we can write the below formula,
\begin{align}
\sum_{m=-\infty}^{\infty} f(m)=\sum_{m=-\infty}^{\infty} \int_{-\infty}^{\infty} f(z) e^{-2 i \pi m z} dz.
\end{align}
Using Eq.~\eqref{final_phase_spacexg0} and applying the above formula , we get
\begin{align}
\label{poisson1}
n_t(x>0,k)&=\frac{1}{N} \sum_{m=-\infty}^{\infty}\!\! T_k \int_0^{\infty} dz\, R_k^{z/N}  \times \\ \nonumber &~~~~~\bigg(\theta(-x-2z+t \sin[k]) \\ \nonumber
&~~~~~-\theta(-x-2 z-2 N+t \sin[k])\bigg) e^{-\frac{2 i \pi m z }{N}}  \\ \nonumber
&= \sum_{m=-\infty}^{\infty} \frac{1}{N} T_k  \int_{\frac{t \sin[k]-x-2 N}{2}}^{\frac{t \sin[k]-x}{2}} R_k^{z/N} e^{-\frac{2 i \pi m z }{N}} dz  .
\end{align}
After performing the integration over $z$ and then extracting the long-time limit by putting $m=0$ in Eq.~\eqref{poisson1}, we get,
\begin{align}
\label{eq:ntxg}
n_t(x>0,k)&=\frac{T_k}{\log[R_k]}\big[R_k^{\frac{-x+t \sin[k]}{2N}}-R_k^{\frac{-x-2N +t \sin[k]}{2N}}\big] \nonumber \\
&=\frac{T_k^2}{R_k |\log[R_k]|} R_k^{-x/2N} R_k ^{t \sin[k]/2N}.
\end{align}
Similarly for $x<0$, using Eq.~\eqref{app_totaln}, we can write the long-time phase space density $n_t(x<0,k)$ as,
\begin{align}
\label{eq:ntxl}
n_t(x<0,k)=\frac{T_k}{|\log[R_k]|}R_k^{\frac{t \sin[k]}{2N}}\big[R_k^{\frac{-|x|}{2N}}+R_k^{-1+\frac{|x|}{2N}}\big].
\end{align}
\noindent
Using both Eq.~\ref{eq:ntxg} and Eq.~\ref{eq:ntxl}, we get an identical expression for absolute value of current $|I|$,
\begin{align}
\label{long_time_current}
|I|=\int_{0}^{\pi} \frac{T_k^2}{R_k |\log[R_k]|} \sin[k] R_k^{\frac{t \sin[k]}{2N}} \frac{dk}{2\pi}.
\end{align}
We now present further analytical results for the case of conformal defect. \\
Conformal defect: For conformal defect, it turns out that the density profile can be computed using Eq.~\eqref{eq:ntxg} and Eq.~\eqref{eq:ntxl} as,
\begin{equation}
\rho(x>0) = 
\frac{\lambda^4 (1-\lambda^2)^{-x/2N} }{2(1-\lambda^2)|\log(
1-\lambda^2)|}  \Big(\pmb{L}_0(A)+I_0(A)\Big)\, ,\\  
\end{equation}
\begin{align}
\rho(x<0) = 
\frac{\lambda^2 \Big[(1-\lambda^2)^{-|x|/2N} + (1-\lambda^2)^{-1+|x|/2N} \Big]}{2(1-\lambda^2)|\log(
1-\lambda^2)|} \nonumber \\
\times \Big(\pmb{L}_0(A)+I_0(A)\Big)\, .
\end{align}
where $\pmb{L}_0(a)$ and $I_0(a)$ are Struve and modified Bessel function of the first kind respectively \cite{olver2010nist}, and the coefficient $A$ is given by   
\begin{equation}
A=\frac{t \log \left(1-\lambda ^2\right)}{2 N}.
\end{equation} 
Using the large argument expansion of Struve and Bessel function, the density profiles for both $x<0$ and $x>0$ can be further simplified to,
\begin{align}
\rho(x>0)&=\frac{\lambda^4 }{\pi|\log[1-\lambda^2]|^2} (1-\lambda^2)^{-\frac{x}{2N}-1} \frac{1}{(t/t_P)}, \nonumber \\ 
\rho(x<0)&=\frac{\lambda^2 }{\pi|\log[1-\lambda^2]|^2} \big[(1-\lambda^2)^{-\frac{|x|}{2N}}  \nonumber\\
&+(1-\lambda^2)^{-1+\frac{|x|}{2N}}\big] 
 \frac{1}{(t/t_P)} .
\end{align}
Therefore, for both $x>0$ and $x<0$, the density at any given point in space decays as $1/t$.

Using the expression of current in Eq.~\eqref{long_time_current}, for conformal defect, we get the analytical expression, 
\begin{align}
\label{asymptotic_current}
I=\frac{ \lambda^4}{\pi(1-\lambda^2)\big|\log[1-\lambda^2]\big|^3}\Bigg(\frac{1}{t/t_P}\Bigg)^2.
\end{align}
Our result in Eq.~\eqref{asymptotic_current}
exactly matches with the current that we get from exact numerics and generalized hydrodynamics with no long time approximation [see Fig.~\ref{fig:number}].
For non-conformal defects, the explicit expressions of current is somewhat cumbersome. However, the very long time limit turns out to be $1/t^{4/3}$, quite different from the $1/t^2$ behaviour for conformal defects seen in Eq.~\eqref{asymptotic_current}. This can be argued by taking small $k$ limit in the integrand of Eq.~\eqref{long_time_current}. For both types of non-conformal defects, at very long times,  Eq.~\eqref{long_time_current} takes the form 
\begin{align}
\label{verylong_time_current}
|I|\approx \frac{F}{t^{4/3} } \int_0^{\infty} dz\, z^3 e^{-Gz^3}\, ,
\end{align}
where $F$ and $G$ are independent of $t$ and depend of details of the non-conformal defects. We have shown in Fig.~\eqref{fig:appendix_other_defect} that for non-conformal defects such as hopping and density defects, current eventually shows $t^{-4/3}$ scaling consistent with Eq.~\eqref{verylong_time_current}.

We now calculate the Yang-Yang hydrodynamic entropy using Eq.~\eqref{eq:ntxg} and Eq.~\eqref{eq:ntxl}. We first consider the case when the hydrodynamic entropy is calculated from the system. For $x<0$, the following approximation holds, 
\begin{align}
n_t(x,k)\log\big(n_t(x,k)\big)+(1-n_t(x,k) \log\big(1-n_t(x,k)\big)\sim  \nonumber\\
n_t \log n_t(x,k)-n_t(x,k),  
\end{align}

We then evaluate Eq.~\eqref{entropydens-hydro} to get $s_{hydro}(x)$. Then using, 
large-argument expansions of Struve and Bessel functions, we can simplify Eq.~\eqref{entropyS-hydro} to get $S_{hydro}^{(S)}$ as 
\begin{align}
\label{page_curve_decay}
S_{hydro}^{(S)} \sim \frac{\lambda^2}{\pi |\log[1-\lambda^2]|^2 (t/t_P)} \int_{-N}^{0} B(x) dx .
\end{align}
With,
\begin{align}
B(x)&=\bigg[(1-\lambda^2)^{-1+\frac{|x|}{2N}}+(1-\lambda^2)^{-\frac{|x|}{2N}}\bigg] \bigg[1-\log[1-\lambda^2]+\nonumber \\ 
&\log\log[1-\lambda^2]-\log[(1-\lambda^2)^{-1+\frac{|x|}{2N}}+(1-\lambda^2)^{-\frac{|x|}{2N}}]\bigg].
\end{align}
Above Eq.~\eqref{page_curve_decay}, clearly explains the long-time Page curve $1/t$ decay in presence of conformal defect. Therefore, the hydrodynamic entropy calculated from the system gives the correct entanglement entropy. For the other non-conformal defects (density and hopping), $S_{hydro}^{(S)}$ shows $t^{-1/3}$ scaling in the long-time limit which we have shown in Fig.~\eqref{fig:appendix_other_defect}. The analytical reasoning for this is similar to the one discussed for the case of current [see Eq.~\eqref{verylong_time_current}].\\

%\mkc{It is to important to note that hydrodynamic entropy calculated from the reservoir for large times deviates from the microscopic von Neumann entanglement entropy. Therefore, a large time analysis of the reservoir entropy $S_{\rm hydro}^{(R)}$ is not of relevance. }

% \mkc{NOT CLEAR 
% Again in a similar way using Eq.~\eqref{eq:ntxg}, we can write,
% \begin{align}
% S_{hydro}^{(R)} \sim \frac{1}{(t/t_P)} \int_{0}^t (1-\lambda^2)^{-\frac{x}{2N}} (C-D x/(2N)) dx .
% \end{align}
% Here, $C$ and $D$ are the constant terms and do not depend on $x$ and $t$. }

\section{Finite size effects: behaviors of density, current, entanglement entropy for other system sizes}
\label{app:finite}
 Here we briefly discuss finite size effects and present results for other system sizes. We have shown in Fig.~\eqref{fig:main_appendix} that all the analysis of density, current, entanglement entropy and hydrodynamic entropy hold for other system sizes and all the features  remain the same.  The only fact is that finite system size $N$ appears in Page time $t_P=2N$ and thus $N$ enters into the physical quantities after the Page time $t_P$.

\bibliography{references}
\end{document}